\documentstyle[aps,manuscript]{revtex}
\begin{document}

\title{Low Temperature Magnetothermodynamics of 
Pr$_{0.7}$Ca$_{0.3}$MnO$_{3}$}

\author{M. Roy$^{1}$, J. F. Mitchell$^{2}$, A. P. Ramirez$^{3}$, 
P. Schiffer$^{1}$$^{,4}$\renewcommand{\thefootnote}{\alph{footnote}}
\footnote{Corresponding author:  schiffer@phys.psu.edu}}

\address{$^{1}$Department of Physics, University of Notre Dame, IN 46556}
 
\address{$^{2}$Material Science Division, Argonne National Laboratory,
 Argonne, IL 60439}

\address{$^{3}$Bell Laboratories, Lucent Technologies, Murray Hill, NJ 07974 }

\address{$^{4}$Dept. of Physics, Pennsylvania State University, 104 Davey Lab, University Park, PA  16802}

\maketitle
\begin{abstract}
We present a detailed magnetothermal study of Pr$_{0.7}$Ca$_{0.3}$MnO$_{3}$, 
a perovskite manganite in which an insulator-metal transition 
can be driven by magnetic field, but also by pressure, 
visible light, x-rays, or high currents.  We find that the 
field-induced transition is associated with a large release of energy which accounts for its strong irreversibility.
In the ferromagnetic metallic state, specific heat and magnetization
measurements indicate a much smaller spin wave stiffness than that
seen in any other ferromagnetic manganite, which we explain in terms
of ferromagnetism among the Pr moments. The Pr ferromagnetism
also appears to influence the low temperature thermodynamic phase diagram
of this material and the uniquely sensitive metastability of the insulating state.

\end{abstract}

\newpage

\section{Introduction}

The rare earth perovskite manganites (R$_{1-x}$A$_{x}$MnO$_{3}$) are associated with a wide variety of fascinating physics due to the strong coupling between their electronic, magnetic and lattice degrees of freedom.  Phenomena observed in these materials include $``$colossal$"$ magnetoresistance, real-space charge ordering, electronic phase separation, and a diverse variety of magnetoelectronic ground states (Ramirez 1997, Coey $et \: al.$ 1999).

Although this entire class of materials displays unusual behavior, one  composition, Pr$_{0.7}$Ca$_{0.3}$MnO$_{3}$, has been shown to display a particularly rich set of phenomena. Upon cooling from high temperatures in low fields, this material undergoes a charge-ordering transition at  T$_{co}$ $\approx$ 230 K   (Jir\'{a}k, $et \: al.$ 1980, Jir\'{a}k $et \: al.$ 1985, Cox $et \: al.$ 1998) and an antiferromagnetic (AFM) transition (at T $\sim$ 150K) (Yoshizawa $et \: al.$ 1995). At lower temperatures (T$_{CAFM}\lesssim$ 110K), the system enters a different state which we will label canted antiferromagnetic (CAFM) based on earlier neutron scattering results, although some recent papers suggest that the CAFM phase is actually an inhomogeneous mixed phase (Frontera $et \: al.$ 2000, Radaelli $et \: al.$ preprint, Deac $et \: al.$ preprint). At lower temperatures (T $\lesssim$ 60 K), application of a field (H $\gtrsim$ 2.5 T), induces a first-order irreversible transition  to a ferromagnetic (FM)) metallic  state, i.e. the system remains in the FM metallic state even after the external field has been removed (although  the charge ordered state is reestablished on subsequent heating above T $\sim$ 60 K (Tomioka $et \: al.$ 1996).  The strong hysteresis associated with this transition is indicated by the hatched region in figure \ref{f1} which  illustrates where the state is history dependent.

Like several other manganite compounds, the resistivity of Pr$_{0.7}$Ca$_{0.3}$MnO$_{3}$ is reduced by orders of magnitude in a magnetic field due to the irreversible field-induced transition (Tomioka $et \: al.$ 1996, Hwang $et \: al.$ 1995, Tomioka $et \: al.$ 1995, Lees $et \: al.$ 1996, Anane $et \: al.$ 1999).  What sets Pr$_{0.7}$Ca$_{0.3}$MnO$_{3}$ apart from the other manganites is that the metastable insulating state also can be driven metallic by electric field (Asamitsu $et \: al.$ 1997, Stankiewicz $et \: al.$ 2000), high pressure (Morimoto $et \: al.$ 1997), exposure to visible light (Miyano $et \: al.$ 1997, Fiebig $et \: al.$ 1998), or x-rays (Kiryukhin $et \: al.$ 1997, Cox $et \: al.$ 1998)).  To characterize the irreversible transition between  these two magnetic states, we have made a detailed study of  both single crystal and polycrystalline  samples of this compound through resistivity, magnetization,  field-dependent specific heat, and magnetocaloric measurements.    We find that there is an enormous release of heat at the field-induced transition at low temperatures, sufficient to raise the sample temperature by a factor of 2, which explains the irreversibility of the transition.  In the FM state at low temperatures, our specific heat and magnetization measurements indicate a ferromagnetic spin wave stiffness which is far below that seen in any other manganite.  The data can most easily be explained as a result of ferromagnetism among the moments associated with the Pr ions, and the results imply that the Pr magnetism may be important to understanding the unusual low temperature properties of this material.  Some of these results have been published previously elsewhere (Roy $et \: al.$ 2000b).
    
\section{Experimental Details}

We studied both a polycrystalline sample of Pr$_{0.7}$Ca$_{0.3}$MnO$_{3}$ synthesized by a standard solid state method 
and a single crystal grown in a floating zone mirror furnace.  Both samples were judged to be 
single-phase based on x-ray diffraction studies, and the results discussed below were qualitatively 
and quantitatively consistent between the two samples (data for the single crystal are shown). 
Resistivity was measured by a standard four probe ac  in-line method and magnetization was 
measured in a Quantum Design SQUID magnetometer. Specific heat was measured by a semi-adiabatic 
heat-pulse method calibrated against a copper standard (Roy 1999, unpublished).

Magnetocaloric measurements were performed using our calorimeter.  The calorimeter with the 
sample attached  was  temperature-controlled at a few degrees above the surrounding cryostat 
temperature after zero-field cooling.  The field was then swept while recording the heat required to 
maintain the sample at  constant temperature.  The difference between the input heat required 
during the field sweep and  that required at constant field was attributed to magnetocaloric effects 
since other sources of heating (e.g. eddy currents) were found to be negligible.

\section{Thermodynamics of the Field-Induced CAFM - FM Transition}

In this section we describe the field dependence of the low temperature thermodynamic properties of Pr$_{0.7}$Ca$_{0.3}$MnO$_{3}$ with particular attention to the field induced CAFM-FM transition at $\sim$  4 T.  The insulating CAFM phase has been shown to have spin-glass-like properties (consistent with phase separation (Cox $et \: al.$ 1998, Yoshizawa $et \: al.$ 1995, Tomioka $et \: al.$ 1995, Frontera $et \: al.$ 2000, Radaelli $et \: al.$ preprint, Deac $et \: al.$ preprint)), and the spin-glass-like nature is evident in the low field temperature dependence of the magnetization, M(T).   At T$_{CAFM}$ we observe a sharp rise in M(T), but at lower temperatures we see a large difference  
between the zero-field-cooled and the field-cooled magnetization for H $\lesssim$ 0.5 T (see figure 
\ref{f2}).  While the magnetization after field-cooling displays an increase with decreasing 
temperature,  magnetization after the sample had been zero-field cooled drops at a lower 
temperature, and continues to decrease with decreasing temperature.  The  extent of this difference 
between the zero-field-cooled and the field-cooled magnetization decreases with increasing field 
with the two curves becoming nearly identical  at fields above 0.5 T.

Upon raising the field after zero-field-cooling, the magnetization, M(H), shows a sharp rise.  The slope of M(H) decreases with increasing field until 
$\sim$ 3 T when there is an increase in dM/dH corresponding to the transition from the 
CAFM state to the FM state (see figure \ref{f3}).  At high fields,  M(T) reaches a value consistent with $\sim$ 95$\%$ of 
full saturation magnetization of the Mn lattice. On subsequent field sweeps, however, M(T) follows 
the FM magnetization curve -- confirming the irreversibility of the first-order field-induced 
transition. The field required to induce the transition decreases with increasing temperature, and at 
T $\gtrsim$ 60 K, M(T) becomes non-hysteretic, indicating the maximum temperature for which 
the FM phase is stable in zero field.

We measured the  field-dependence of the specific heat, C(H), by zero-field cooling the sample and 
then measuring the specific heat every 0.25 T while sweeping the field from 0 $\rightarrow$ 9 T 
$\rightarrow$ -9 T $\rightarrow$ 9 T.  We find that C(H) decreases monotonically with increasing 
field which is consistent with suppression of spin excitations in a highly magnetized system (see figure 
\ref{f4}).  Upon raising the field after zero-field-cooling, there is  a sharp drop in C(H) at H $\sim$ 
4 T corresponding to the CAFM-FM transition.    The large difference between the specific heats of 
the CAFM and the FM states reflects the first order nature of the transition, and is plotted at H = 0 
in the inset to figure \ref{f4}.  Even within the FM state, the specific heat is 
extraordinarily sensitive to magnetic field, changing  by $\sim$  70 mJ/mole-K or  $\sim$ 40$\%$ between 0 and 9 T.  
The magnitude of this change, and its implications for the thermodynamics of Pr$_{0.7}$Ca$_{0.3}$MnO$_{3}$ will be 
discussed below.

The sample mounted on the calorimeter is only weakly thermally linked to the surrounding 
cryostat, and in the course of measuring C(H) it became evident that a significant amount of heat 
was being released by the sample upon passing through the CAFM-FM transition.  This 
self-heating is illustrated in figure \ref{f5} which shows the temperature of the sample (T$_{S}$) 
during a field sweep.   For this measurement, the sample was  cooled in zero field and stabilized at the temperature of the surrounding cryostat, then the sample  temperature  was measured as the  field was swept from 0  $\rightarrow$ 9 T, 9 T $\rightarrow$ 0 and  0 $\rightarrow$ 9 T at 0.0006 T/second.  As seen in 
figure \ref{f5}, when the field is raised from 0 to 9 T for the first time, T$_{S}$  increases 
remarkably with increasing field, displaying a steady rise  between around 2.5 T to 4 T. At higher 
fields  (4 T $\lesssim$ H $\lesssim$ 5.5T), while dT$_{S}$/dH is negative, the large temperature 
difference between the sample and the base temperature indicates continued self-heating.  This was 
confirmed by stopping a field sweep at 4T, allowing the sample to equilibrate to the surrounding 
cryostat temperature, and then observing a rise in  T$_{S}$ upon resuming the field sweep. The 
sample continues to show some heating  effect even at higher fields, although not as pronounced as 
that at low fields. On decreasing the field and during subsequent field sweeps,  T$_{S}$ remains 
largely constant except for small changes attributable to demagnetization effects.

To quantify this self-heating, we measured the actual heat released by the sample through 
magnetocaloric measurements.  These measurements were conducted by monitoring the input 
power (P) required to maintain the calorimeter at a constant temperature 
during field sweeps. Figure \ref{f6} shows a typical dataset at low temperatures in which we first 
cooled in zero field and then swept the magnetic field 0 $\rightarrow$ 9 T $\rightarrow$ -9 T 
$\rightarrow$ 9 T while keeping both the calorimeter and the surrounding cryostat at constant 
temperature. When the field is raised for the first time after zero field cooling, we see a series of 
small features in P(H) for H $<$ 1 T.   We attribute these reproducible features to heat release 
associated with the spin-glass-like character of the zero-field-cooled CAFM state  
(Yoshizawa $et \: al.$ 1995, Anane $et \: al.$ 1999, Deac $et \: al.$ preprint), i.e. irreversible relaxation of the  spin configuration during the initial field 
sweep (Mydosh 1993, Tsui $et \: al.$ 1999).  These low field features are not seen in P(H) in the FM phase after 
field cooling or after a large field has been applied and removed, although there is a slight drop near H = 0 which we attribute to domain effects.  The most dramatic feature in the magnetocaloric data, however, is the heating at higher fields (H $\gtrsim$ 2.5 T) during the initial sweep up in field, as reflected by the 
negative peak in P(H).  This peak is associated with the first order CAFM-FM phase transition and, 
on subsequent field sweeps, the sample displays essentially  no heating, demonstrating the 
irreversible nature of this heat  release.  We calculated the total heat released in this process 
between  H $=$ 2.5T and 9 T using

\begin{equation}
{Q = \int\limits_{2.5 T}^{9 T}\frac{P(H)}{dH/dt}dH}.
\label{e1}
\end{equation}
Since the sweep rate (dH/dt) is constant, the total heat released by the sample can be evaluated quite easily by using 
equation \ref{e1} and subtracting the background (using the data from subsequent field sweeps as 
the baseline). The total heat released in this  process is found to be around 15 $\pm$ 1 J/mole which 
is enormous -- large enough to increase the sample temperature of a perfectly thermally 
isolated sample  from 5 to 15 K. Moreover, this  heating is not an artifact of eddy current  heating, 
since Q remains  unchanged (within a  few percent) even when the sweep rate is  increased by a 
factor of 4 (see figure  \ref{f7}). Similar studies were performed at different temperatures, and we 
find that Q is only weakly dependent on temperature at low temperatures (T $\lesssim$ 30 K).  
At higher temperatures, Q decreases monotonically with increasing temperature above T $\sim$ 30 
K, before finally disappearing at T $\sim$ 50 K as shown in figure  \ref{f8}.

        The magnitude of Q from the magnetocaloric measurements is extraordinarily large 
compared with the other thermal energy scales in the system, as evidenced by the large 
self-heating.  The 15 J/mole obtained at low temperatures is, in fact, significantly larger than what 
one would calculate for the free energy based on an integration of the specific heat.  On the other hand, the source of this large amount of heat is readily apparent from our measurements of M(H) at low temperatures.  The increment in magnetization associated with the CAFM-FM transition occurs in a magnetic field, and therefore must be associated with a 
decrease in the Zeeman energy of the sample, the release of which should be observed as heat.  One 
can calculate this magnetic heat  release Q$_{M}$ from simple thermodynamics using the 
following integral:

\begin{equation}
{Q_{M} = \int\limits^{H_{max}}_{H_{min}}HdM
\longrightarrow Q_{M} = 
\int\limits^{H_{max}}_{H_{min}}H\frac{\partial{M}}{\partial{H}}dH},
\label{e2}
\end{equation}
where H$_{min}$ and H$_{max}$ are defined by the width of the transition. This integral is readily evaluated  
from our measurements of M(H), and, at low temperatures, we find Q$_{M}$ $\sim$ 10 $\pm$ 1 
J/mole with error bars calculated from the uncertainty in determining the limits of the integral.  
This magnitude is consistent with the magnetocaloric measurements  discussed above, confirming 
that this is the source of the large heat release at the transition.  We find that Q$_{M}$ decreases  
with increasing temperature (see figure \ref{f8}),  consistent with the disappearance of this field 
induced transition  at T $\gtrsim$ 60 K.   The thermodynamic origin of this large heat release is 
discussed below in context of ferromagnetism among the Pr ions.

\section{Thermodynamics of the Ferromagnetic Phase}

        One of the most striking features of the data discussed above is the rather strong field 
dependence of the low temperature specific heat in the FM phase (i.e. after the material undergoes 
the irreversible field  induced transition).  The  absolute value of the drop in C(H) between 0 and 9 
T in the FM  state  is at least a factor of 15 larger than that of  the other FM conducting manganites 
such as  La$_{0.7}$Ca$_{0.3}$MnO$_{3}$ or  La$_{0.7}$Sr$_{0.3}$MnO$_{3}$ (Roy $et \: al.$ 2000a).  
In order to understand the origin of this large drop in the specific heat, we also measured the 
specific heat as a function of temperature at various fields. In order to enter the FM state, we either 
field-cooled the sample (for H $>$ 4T) or raised the field to 9 T at low temperatures and then 
reduced it to the desired value (for H $<$ 4T).     As shown in figure \ref{f9}, we find that C(T) in 
the FM state is well described by a combination of phonon and spin wave terms:

\begin{equation}
{C(T) = \beta T^{3} + \delta T^{3/2}}
\label{e3}
\end{equation}
where the first term corresponds to the lattice specific heat and the second corresponds to the 
specific heat of FM magnons at H = 0.  This choice of fitting function can be justified by subtracting the lattice contribution ($\beta T^{3}$) from the raw data.  The result follows a power law temperature dependence with an exponent of 1.5 as seen in the inset to figure \ref{f9}, confirming that the remainder is the magnetic contribution C$_{mag}$ $=$ $\delta T^{3/2}$.   This form is appropriate for H = 0, and it fits the zero-field data well.  That it also appears to fit the data at higher fields, even though the corrections for field dependence have not been included, may be attributable to some temperature dependence in the spin stiffness which would be reflected in $\delta$. It should also be noted, that while this form fits the data well, we cannot rule out the possibility of some additional contribution, such as a linear term which would arise from free electrons or a spin glass (Smolyaninova $et \: al.$ 2000).  Given the quality of the fit and also the fits to C(H) discussed below, however, we expect such a contribution to be negligible.

>From our fits to C(T) we find that $\beta$ remains independent of H (within experimental uncertainty), but that 
$\delta$ decreases rapidly with increasing field (see figure \ref{f10}).  This implies that the 
field dependence of the specific heat at low temperatures can be attributed to the spin-wave 
contribution and the decrease in specific heat with increasing field can be attributed to the field 
suppression of the spin-waves.  We can fit the C(H) data to the form expected for a Heisenberg 
ferromagnet (Kittel 1964):

\begin{equation}
{C(H) = A+\frac{k_{B}^{5/2}T^{3/2}V_{mole}}{4\pi^{2}D^{3/2}}
\int\limits_{g\mu_{B}H/k_{B}T}^{\infty}
\frac{x^2 e^{x}}{{({e^{x}-1})}^2}{\sqrt{x-\frac{g\mu_{B}H}{k_{B}T}}}dx}, 
\label{e4}
\end{equation}
where $V_{mole}$ is the molar volume and the only two fitting constants A and D parameterize the lattice contribution 
and the stiffness constant of the spin wave spectrum respectively.  Such a fit is illustrated by the 
solid line in \ref{f4}.  We fit for $|$H$|$ $>$ 1 T to avoid domain effects which are evident also in our magnetocaloric and magnetization data described above.  The fits to 
C(H) at T = 5.5K yield a stiffness constant (D $\sim$ 28.0 $\pm$ 0.3 meV-{\AA}$^{2}$) which is at least a 
factor of 4 smaller than that of other ferromagnetic metallic manganites (see table \ref{t4.1} (Martin $et \: al.$ 1996, Perring $et \: al.$ 1996, Fernandez-Baca $et \: al.$ 1998, Lynn $et \: al.$ 1996), 
accounting for the much larger suppression of specific heat by an applied field. A fit at T = 12 K, is of similarly high quality, and yields D = $\sim$ 22.0 $\pm$ 0.5 meV-{\AA}$^{2}$, suggesting a slight temperature dependence to the spin stiffness.
        The very soft spin wave spectrum implied by the fits to C(H) is quite surprising since there 
should be little difference in the Mn-Mn exchange energies in Pr$_{0.7}$Ca$_{0.3}$MnO$_{3}$ and other manganites which have similar structures and lattice constants.  The 
value of D can be checked, however, from the fits to the temperature dependence of the specific 
heat C(T). The fit parameter $\delta$ for H $=$ 0 is related to the spin stiffness constant through 
$C_{magnon} = 0.113k_{B}V_{mole}\left(\frac{k_{B}}{D}\right)^{3/2}$, and thus we can calculate the 
stiffness constant using the value of $\delta$ from the fit to FM state C(T) at H = 0. We find that 
this value of D (25.3 $\pm$ 0.5 meV-{\AA}$^{2}$) is close to that calculated from the 
C(H) data.

To further test the extraordinarily low value of D obtained from specific heat data, we measured the 
low temperature field-cooled temperature dependence of the magnetization in the FM state at H 
$=$ 7 T. When proper corrections are made for the field-induced spin gap in an  external field, 
M(T) is consistent with spin-wave excitations with the exception of the lowest temperature data 
which is suppressed  by around 0.3$\%$ possibly due to a minute presence of a second phase as 
shown in figure \ref {f11}. Modeling the data as a simple Heisenberg ferromagnet and assuming 
that there is no spin gap  at H $=$ 0, the spin-wave dispersion relation at an external field H is

\begin{equation}
{\omega_{k} = g \mu _{B}(H-NM)+Dk^{2}},        
\label{e5}
\end{equation}
where NM is the demagnetization field and M is the magnetization. Using the standard spin-wave 
picture, the magnetization is given by (Kittel 1964, Kunzler $et \: al.$ 1960, Henderson $et \: al.$ 1969, Smolyaninova $et \:al.$ 1997):

\begin{equation}
{M(0,H)-M(T,H)={g\mu_{B}}\left(\frac{k_{B}T}{4 \pi D}\right)^{3/2} f_{3/2}(g \mu _{B}(H-
NM)/k_{B}T)}, 
\label{e6}
\end{equation}
where
\begin{equation}
{f_{p}(y) = \sum_{n=1}^{\infty} \frac{e^{-ny}}{n^{p}}}.
\label{e7}
\end{equation}
Our sample had dimensions 1.04mm$\times$0.904mm$\times$0.4mm, and thus the 
demagnetization field at 7 T was calculated to be around 0.54 T. The summation f$_{3/2}(y)$  in 
equation \ref{e7} is evaluated at discrete temperatures, and the  stiffness constant, D is calculated 
from a plot of  $(M(0,H)-M(T,H))/f_{3/2}T^{3/2}$ as shown in the inset to figure  \ref{f11}. We 
find that the measured value  of D of this compound is around 25 $\pm$ 2 meV-{\AA}$^{2}$,  
this is again at least a factor of four smaller than that of other ferromagnetic metallic perovskite manganites (see table 
\ref{t4.1}), and consistent with the specific heat data.

\section{Discussion}

Since the fits to C(T), C(H) and M(T) consistently yield a spin stiffness constant which is 
much smaller than that obtained in other manganites, we must look for an explanation as to what 
might be different in Pr$_{0.7}$Ca$_{0.3}$MnO$_{3}$.  To answer this question, we note that the FM saturation 
magnetization of Pr$_{0.7}$Ca$_{0.3}$MnO$_{3}$ is at  least 10$\%$ higher than that in other FM metallic rare-earth manganites such 
as La$_{1-x}$Ca$_{x}$MnO$_{3}$ (see table \ref{tmag}) (Tomioka $et \:al.$ 1995, Lees $et \: al.$ 1996, Thomas $et \: al.$ 1999, Gong $et \: al.$ 1995, Urushibara $et \: al.$ 1995, Ju $et \: al.$ 1995, Martinez $et \: al.$ 1996).  Since 
the Mn spins display considerable canting (Yoshizawa $et \: al.$ 1995) in Pr$_{0.7}$Ca$_{0.3}$MnO$_{3}$  even at high fields, an 
additional FM moment such as that associated with the rare-earth is required to explain this  excess 
magnetization. Recent neutron diffraction studies (Cox $et \: al.$ 1998) have indeed observed 
ferromagnetic  ordering of the Pr$^{3+}$ ions for T $<$  60 K.   Moreover, we also observe a 
distinct small rise in M(T) at T $\sim$ 50-60K in figure 2 with an accompanying peak in C(T), 
presumably associated with the weak  FM ordering of the Pr moments.  Based on the similarity in the lattices, there is no reason to 
expect that the Mn-Mn FM interaction is more than 4 times weaker in Pr$_{0.7}$Ca$_{0.3}$MnO$_{3}$ than that in other FM 
manganites, and in fact preliminary neutron scattering measurements of D for the Mn spins in the field-induced ferromagnetic phase of
Pr$_{0.7}$Ca$_{0.3}$MnO$_{3}$ yield D $\sim$ 150 meV-{\AA}$^{2}$ (Fernandez-Baca and Dai, unpublished). We therefore hypothesize that the weak 
FM ordering of the Pr ions is responsible for small spin stiffness in Pr$_{0.7}$Ca$_{0.3}$MnO$_{3}$.  This hypothesis implies that 
excitations among the Pr spins dominate the low temperature thermodynamics of Pr$_{0.7}$Ca$_{0.3}$MnO$_{3}$ since the 
softer spin wave spectrum of the Pr moments is more sensitive to thermal excitations than that of 
the Mn spins.

        We now discuss the origin of the large heat release at the first order CAFM-FM transition, 
which, at first consideration, might be considered as the latent heat of the transition.  If a system 
undergoing a first order phase transition is in equilibrium throughout the transition, then the free energy changes continuously through the transition.  The latent heat associated with such a transition results from a 
discontinuity in the first temperature derivative of the free energy, i.e.  a discontinuous change in 
the entropy of the system even though the system is always in equilibrium.  In addition to the latent 
heat, more energy can be released at a first order transition if the system is not in equilibrium when 
the transition occurs,  e.g. in the case of supercooling or supermagnetization.  In such cases, there  
is an energetic barrier to the nucleation of the equilibrium phase which is typically characterized as a 
positive surface energy of the interface between the two phases.  Homogeneous nucleation of the 
equilibrium phase only occurs when the system is driven so far out of equilibrium that the free energy difference between the two phases becomes 
large enough for thermal fluctuations to overcome this surface energy barrier.  Once the equilibrium phase nucleates, the excess free energy is typically released as heat.

In the case of Pr$_{0.7}$Ca$_{0.3}$MnO$_{3}$ cooled in zero field, the CAFM state seems to be the lowest energy state for T $\gtrsim$ 60 K, the same temperature at which the Pr moments order ferromagnetically.  Below 60 K, the FM state apparently becomes energetically more favorable than the CAFM state even at H = 0 since application of x-rays induces metallicity which is then quenched upon raising the field above 60 K. Thus for T $\lesssim$ 60 K, the CAFM state resulting from zero-field cooling is metastable and effectively strongly supercooled.  At this low temperature, however, the energetic advantage of the FM state is apparently insufficient to cause it to nucleate spontaneously through thermal fluctuations, presumably due to a relatively large effective surface energy, and thus the system remains in the non-equilibrium CAFM state.  A large applied field can, however, make the FM state more energetically favorable, and in a large enough field the non-equilibrium free energy difference between the two phases becomes sufficient to allow homogeneous nucleation of the FM phase.  When the phase transition is induced, this free energy difference associated with the applied field is released as heat, which is what we observe in our magnetocaloric measurements.  Once the system is in the FM state, there is no equivalent energetic advantage to reverting to the CAFM state when the field is removed, and thus the transition is irreversible.  This scenario for the low temperature zero-field thermodynamics is shown schematically in figure \ref {f12}.

Within this scenario we can also understand the reversible nature of the light-induced 
insulator-metal transition in Pr$_{0.7}$Ca$_{0.3}$MnO$_{3}$.  When light is applied to Pr$_{0.7}$Ca$_{0.3}$MnO$_{3}$ at low temperatures, it drives small 
regions of the sample into the FM conducting state.  When the light 
and applied voltage are removed, however, the sample returns to the CAFM insulating state (Miyano $et \: al.$ 1997, Fiebig $et \: al.$ 1998).  Based on the 
stability of the low temperature CAFM phase after cooling in zero field, one expects 
that the surface energy at the interface between the CAFM and FM states is large.  Thus the small regions of 
the FM conducting phase will be "swallowed" by the CAFM phase when the external stimuli 
are removed, since the system will gain more energy by shrinking the interface between the two 
phases than by growing the FM phase from these small regions.

\section{Conclusions}

We have presented a variety of thermodynamic data on Pr$_{0.7}$Ca$_{0.3}$MnO$_{3}$ including magnetization, 
magnetocaloric and specific heat measurements, both as a function of temperature and applied 
magnetic field.  Our data suggest that the Pr ferromagnetism and its coupling with Mn spins may be important to the physics of this system at low temperature, and that excitations among the Pr spins 
might dominate the thermal excitations in the material at low temperatures.  The Pr ordering temperature (T $\sim$ 60)  is coincident with the maximum temperature at which  FM phase is stable at H = 0 (induced either by pressure or a previously applied field).  Furthermore, this is also the temperature above  which the x-ray induced metallicity is quenched (Kiryukhin $et \: al.$ 1997).  It thus appears possible that the onset of Pr ferromagnetism destabilizes the CAFM phase relative to the FM phase, so that in zero field the CAFM phase is stable for T $>$ 60 K but then becomes unstable to the FM phase below T $\sim$ 60 K when the Pr spins order.   In this scenario, both the Pr ferromagnetism and its coupling to the Mn moments  are crucial to understanding the physics 
and are therefore inseparable from the numerous unique phenomena 
observed in Pr$_{0.7}$Ca$_{0.3}$MnO$_{3}$.


\begin{figure}
\caption {The low temperature field-temperature phase diagram of
Pr$_{0.7}$Ca$_{0.3}$MnO$_{3}$ where the shaded region indicates 
the history dependent region (reproduced from~\protect(Tomioka $et \: al.$ 1996)).  }
\label{4phase}
\label{f1}
\end{figure}

\begin{figure}
\caption {The temperature dependence of the magnetization at low fields after field cooling (closed 
circles) and zero-field cooling   (open circles). The solid line is a guide to the eye to emphasize the 
low temperature feature in M(T) corresponding to FM ordering of the Pr ions. }
\label{f2}
\end {figure}

\begin{figure}
\caption {The field dependent magnetization illustrating the field induced irreversible transition 
from a canted AFM to FM state at T $=$ 10 K.  The measurements are done when the field is swept 
from 0 $\rightarrow$ 7 T  (solid line), 7 T $\rightarrow$ -7 T (open circles), -7 T $\rightarrow$ 7 T  
(dashed line).  The inset shows the derivative dM/dH during the initial sweep up in field after zero-
field cooling, where the maximum indicates the field-induced transition.}
 \label{f3}
 \end {figure}

\begin{figure}
\caption {The zero field cooled field dependent specific heat at T = 5.5 K, when the field is swept 
from 0 $\rightarrow$ 9 T (solid circles), 9 T $\rightarrow$ -9 T (open triangles), and -9 T 
$\rightarrow$ 9 T (solid circles). The  solid lines are the fit to the data as explained in the text. The 
inset shows temperature dependence of the hysteresis ($\Delta$C) in C(H), i.e.,  the difference in 
zero-field-cooled C(H) at H = 0 and C(H) at H = 0 after the field is raised to 9 T and subsequently 
dropped to 0.   } \label{f4}
\end {figure} 
    
\begin{figure}
\caption {Change in temperature of the sample on the thermally isolated calorimeter during field 
sweeps, indicating the large self-heating of the Pr$_{0.7}$Ca$_{0.3}$MnO$_{3}$ sample.  The data are taken when the  field is 
changed from H = 0 $\rightarrow$ 9 T (solid line), H = 9 T $\rightarrow$ -9 T (dashed line)  and H 
= -9 T $\rightarrow$ 9 T  (dotted line) at different rates as indicated in the figure. }
\label{f5}
\end {figure}

\begin{figure}
\caption { Heat released by the sample, when it is zero-field-cooled and the sample temperature is 
maintained at T $=$ 9.300 $\pm$ 0.025 K, and the  surrounding cryostat temperature is at 7.5 K. 
The data is taken when the  field is changed from H = 0 $\rightarrow$ 9 T (solid line),  H = 9 T 
$\rightarrow$ -9 T (dashed line) and H = -9 T $\rightarrow$ 9 T  (dotted line) at the rate of 6 
gauss/sec. The inset shows a magnification of the low field portion of the data}
\label{f6}
\end {figure}

\begin{figure}
\caption { Sweep rate dependence of the magnetocaloric measurement of the power P(H) required 
to maintain the sample at T $=$ 9.300 $\pm$ 0.025 K while the  surrounding cryostat temperature 
is at 7.5 K. The data are taken after zero-field-cooling when the field is changed from H = 0 
$\rightarrow$ 9 T (solid line),  H = 9 T $\rightarrow$ -9 T (dashed line) and H = -9 T $\rightarrow$ 
9 T  (dotted line) at the rates of 6,12, and 24 gauss/sec.}
\label{f7}
\end {figure}

\begin{figure}
\caption { The heat-release, Q$_{M}$, calculation from the reduction of Zeeman energy associated 
with the CAFM -FM transition.}
\label{f8}
\end {figure}

\begin{figure}
\caption {The zero field cooled specific heat in the ferromagnetic state (as discussed in the text) as 
a function of temperature at H = 0 (open circles), H = 3 T (open down-triangles), H = 6 T (open up-
triangles) and H = 9 T ((open squares). The solid lines are the fit to the  data as discussed in the 
text. The inset shows the C$_{mag}$ vs. T in the  FM state at H = 0 and H = 9 T on a log-log 
scale and the solid  line is the best linear fit with a slope of 1.5, the dashed and the dotted lines 
have slopes of 1 and 2 respectively.  (Reproduced from Roy $et \: al.$ (2000b)).} 
\label{f9}
\end {figure}

\begin{figure}
\caption {The specific heat fitting parameters $\beta$ (top panel) and $\delta$ (bottom panel)
as a function of field. } 
\label{f10}
\end {figure}

\begin{figure}
\caption {The field cooled magnetization at H = 7 T. The solid line is a fit to the data as described in the text. The inset
illustrates weighted-fit for the calculation of D as discussed in the text~\protect(Reproduced from Roy $et \: al.$ (2000b)). }
\label{f11}
\end {figure}

\begin{figure}
\caption {A schematic of the hypothesized zero-field magnetothermodynamics in  Pr$_{0.7}$Ca$_{0.3}$MnO$_{3}$.  a. At temperatures below the CAFM ordering temperature of the Mn spins and above the ordering temperature of the Pr moments, the Mn spins are in their equilibrium CAFM configuration and the Pr moments are disordered.  b.  After zero-field-cooling to below the ferromagnetic ordering temperature of the Pr moments, the Mn spins retain the CAFM order although they are not in their equilibrium state due to the coupling with the Pr moments (as indicated by the dashed arrows).  c. After raising and lowering the magnetic field at low temperatures, the Mn moments enter the equilibrium FM state. }
\label{f12}
\end {figure}

\begin {table}[h] 
\begin{center}
\caption{Stiffness Constant of Ferromagnetic Rare-Earth Manganites.}
\begin {tabular}{cccc} 
Sample &  Ref  &  T$_{c}$(K)  &  D(0)(meV-{\AA}$^{2}$) \\
\tableline
La$_{0.70}$Sr$_{0.30}$MnO$_{3}$  &  Martin $et \: al.$ 1996     & 378.0 & 188\\ 
La$_{0.30}$Pb$_{0.30}$MnO$_{3}$  &  Perring $et \: al.$ 1996     & 355.0 & 134\\
Pr$_{0.63}$Sr$_{0.37}$MnO$_{3}$   &  Fernandez-Baca $et \: al.$ 1998     & 300.9 & 165\\
Nd$_{0.70}$Sr$_{0.30}$MnO$_{3}$  &   Fernandez-Baca $et \: al.$ 1998    & 197.9 & 165\\
La$_{0.67}$Ca$_{0.33}$MnO$_{3}$  &  Lynn $et \: al.$ 1996     & 250.0 & 170\\
\label{t4.1}
\end {tabular}
\end{center}
\end{table}

\begin {table}[h] 
\begin{center}
\caption{Low Temperature Saturation moment of Ferromagnetic 
Rare-Earth Manganites.}
\begin {tabular}{cccc}  
Sample &  Ref  &   Saturation moment (emu/mole) \\
\tableline
La$_{1-x}$Sr$_{x}$MnO$_{3}$ (x $\sim$ 1/3)  &  Hwang $et \: al.$ 1995, Urushibara $et \: al.$ 1995  &  18000-22000\\ 
La$_{1-x}$Ca$_{x}$MnO$_{3}$  (x $\sim$ 1/3)  &  Hwang $et \: al.$ 1995, Gong $et \: al.$ 1995  &  
20000-215000\\
La$_{0.30}$Ba$_{0.30}$MnO$_{3}$  &  Ju $et \: al.$ 1995  &  20000\\
(La$_{0.93}Y_{0.07})_{0.67}$Ca$_{0.33}$MnO$_{3}$  &  Mart\'{i}nez $et \: al.$ 1996  &  
20000\\
Pr$_{0.70}$Ca$_{0.30}$MnO$_{3}$  &  Lees $et \: al.$ 1996  &  23495\\
Pr$_{0.70}$Ca$_{0.30}$MnO$_{3}$  &  Tomioka $et \: al.$ 1995  &  23450\\
\label{tmag}
\end {tabular}
\end{center}
\end{table}

\begin{center}
{\bf REFERENCES}
\end {center}
\begin {description}
\item[]
Anane, A., Renard, J. -P., Reversat, L., Dupas, C., Veillet, P., Viret, M., Pinsard, L. and Revocolevschi, A. 1999, $Phys. \: Rev.$ {\bf B59}, 77.
\item[]
Asamitsu, A., Tomioka, Y., Kuwahara, H. and Tokura, Y. 1997, $Nature$ {\bf 388}, 50. 
\item[]
Coey, J. M. D., Viret, M. and von Molnar, S., 1999 $Adv. \: Physics$ {\bf 48},167.
\item []
Cox, D. E., Radaelli, P. G., Marezio, M. and Cheong, S. -W., 1998, $Phys. \: Rev.$ {\bf B57}, 3305.

\item[]
Deac, I. G., Mitchell, J. F., Schiffer, P. (preprint, cond-mat/0012143)
\item[]
Fernandez-Baca, J. A., Dai, P., Hwang, H. Y., Kloc, C. and Cheong, S. -W. 1998, $Phys. \: Rev. \: Lett.$ {\bf 80}, 4012.
\item[]
Fernandez-Baca, J. A. and Dai, P. (unpublished).
\item []
Fiebig, M., Miyano, K., Tomioka, Y. and Tokura, Y., 1998, $Science$ {\bf 280}, 12.
\item[]
Frontera, C., Garcia-Munoz, J. L., Llobet, A., Respaud, M., Broto, J. M., Lord, J. S. and Planes, A. 2000, $Phys. \: Rev.$ {\bf B62}, 3381.
\item[]
Gong, G. Q., Canedy, C., Xiao, G., Sun, J. Z., Gupta, A. and Gallagher, W. J., 1995, $Appl. \: Phys. \: Lett.$ {\bf 67}, 1783.
\item []
Henderson, A. J., Onn, D. G., Meyer, H. and Remeika, J. P., 1969, $Phys. \: Rev.$ {\bf 185}, 1218. 
\item[]
Hwang, H. Y., Cheong, S. -W., Radaelli, P. G., Marezio M. and Batlogg, B., 1995, $Phys. \: Rev. \: Lett.$ {\bf 75}, 914.
\item[]
Jir\'{a}k, Z., Krupi\v{c}ka, S., Nekvasil, V., Pollert, E., Villeneuve, G. and Zounov\'{a}, F., 1980, $J. \: Mag. \: Mag. \: Mater.$ {\bf 15-18}, 519. 
\item []
Jir\'{a}k, Z., Krup\v{i}cka, S., Simsa, Z., Dlouha, M. and Vratislav, Z., 1985, $J. \: Mag. \: Mag. \: Mater.$ {\bf 53}, 153.  
\item[]
Ju, H. L., Gopalkrishnan, J., Peng, J. L., Li, Q., Xiong, G. C., Venkatesan, T. and Greene, R. L., 1995 $Phys. \: Rev.$ {\bf B51}, 6143.
\item[]
Kiryukhin, V., Casa, D., Hill, J. P, Keimer, B., Vigliante, A., Tomioka Y. and Tokura, Y., 1997, $Nature$ {\bf 386}, 813. 
\item[]
Kittel, C., 1964 {\em Quantun Theory of Solids} (New York: John Wiley and Sons, Inc.).
\item[]
Kunzler, J. E., Walker L. R and Galt, J. K., 1960, $Phys. \: Rev.$ {\bf 119}, 1609. 
\item[]
Lees, M. R., Barratt, J., Balakrishnan, G., McK Paul D. and Dewhurst, C. D., 1996, $J. \: Phys. \: Condens. \: Matter.$ {\bf 8}, 2967. 
\item[]
Lynn, J. W., Erwin, R. W., Borchers, J. A., Huang, Q. and Santoro, A., 1996, $Phys. \: Rev. \: Lett.$ {\bf 76}, 4046.
\item []
Martin, M., Shirane, G., Endoh, Y., Hirota, K., Moritomo, Y. and Tokura, Y., 1996, $Phys. \: Rev.$ {\bf B53}, R14285. 
\item []
Mart\'{i}nez, B., Fontcuberta, J., Seffar, A., Garc\'{i}a-Mu\~{n}oz, J. L., Pi\~{n}ol, S. and Obradors, X., 1996, $Phys. \: Rev.$ {\bf B54}, 10001.
\item []
Miyano, K., Tanaka, T., Tomioka, Y. and Tokura, Y., 1997, $Phys. \: Rev. \: Lett.$ {\bf 78}, 4257. 
\item[]
Morimoto, Y., Kuwahara, H., Tomioka Y. and Tokura, Y., 1997, $Phys. \: Rev.$ {\bf B 55}, 7549. 
\item[]
Mydosh, J. A., 1993 {\em Spin Glasses: An Experimental Introduction}, (London:  Taylor and Francis). 
\item[]
Perring, T. G., Aeppli, G., Hayden, S. M., Carter, S. A., Remeika, J. P. and Cheong, S. -W., 1996 $Phys. \: Rev. \: Lett.$ {\bf 77}, 711.
\item[]
Radaelli, P. G., Iberson, R. M., Cheong, S. -W. and Mitchell, J. F. (preprint,cond-mat/0006190).
\item[]
Ramirez, A. P., 1997 $J. \: Phys. \: Condens. \: Matter$ {\bf 9}, 8171. 
\item[]
Roy, M., Mitchell, J. F., Potashnik, S. J. and Schiffer, P., 2000a, $J. \: Mag. \: Mag. \: Mat.$, {\bf 218}, 191.
\item[]
Roy, M., Mitchell, J. F., Potashnik, S. J. and Schiffer, P., 2000b, $Phys. \: Rev. \: B$.
\item []
Roy, M., 1999, Ph.D. Thesis, University of Notre Dame (unpublished).
\item[]
Smolyaninova, V. N., Biswas, A., Zhang, X., Kim, K. H., Kim, B. -G., Cheong, S. -W. and Greene, R. L., 2000, $Phys.\: Rev.$ {\bf B 62}, 6093. 
\item[]
Smolyaninova, V. N., Hamilton, J. J., Greene, R. L., Mukovskii Y. M. and Karabashev, S. G.  and Balbashov, A. M., 1997, $Phys. \: Rev.$ {\bf B 55}, 5640. 
\item[]
Stankiewicz, J., Sese, J., Garcia, J., Blasco, J. and Rillo, C. 2000 $Phys. \: Rev.$ {\bf B61}, 11236.
\item[]
Thomas, R. -M., Skumryev, V., Coey, J. M. D. and Wirth, S., 1999, $J. \: Appl. \: Phys.$ {\bf 85}, 5384.
\item[]
Tomioka, Y., Asamitsu, A., Kuwahara, H., Moritomo, Y. and Tokura, Y., 1996, $Phys. \: Rev.$ {\bf B53}, R1689. 
\item[]
Tomioka, Y., Asamitsu, A., Moritomo, Y. and Tokura, Y., 1995, $J. \: Phys. \: Soc. \: of \: Japan.$ {\bf 64}, 3626.
\item[]
Tsui, Y. K., Burns, C. A., Snyder, J. and Schiffer, P., 1999, $Phys. \: Rev. \: Lett.$ {\bf 82}, 3532. 
\item[]
Urushibara, A., Moritomo, Y., Arima, T., Asamitsu, A., Kido, G. and Tokura, Y., 1995, $Phys. \: Rev.$ {\bf B51}, 14103.
\item[]
Yoshizawa, H., Kawano, H., Tomioka, Y. and Tokura, Y., 1995, $Phys. \: Rev.$ {\bf B 52}, R13145.

\end{description}

\end{document}